\documentclass[final]{svjour2}
\usepackage{graphicx}
\usepackage{rotating}
\usepackage{amssymb}
\usepackage{mathptmx}
\usepackage[numbers]{natbib}
\makeatletter
\journalname{Journal of Low Temperature Physics}

\bibpunct{}{}{,}{s}{}{,}

\begin{document}

\newcommand{\hdblarrow}{H\makebox[0.9ex][l]{$\downdownarrows$}-}
\title{Effect of Rough Walls on Transport in Mesoscopic $^3$He Films}

\author{P. Sharma$^1$ \and A. Corcoles$^2$ \and R. G. Bennett $^3$\and J. M. Parpia$^3$ \and B. Cowan$^1$ \and J. Saunders$^1$}

\institute{1:Department of Physics, Royal Holloway University of London, Egham, TW20 0EX, \\Surrey, UK\\
\email{priya.sharma@rhul.ac.uk}
\\2: Present address, IBM Watson Research Center, Yorktown Heights, NY 10598, USA\\
3: Department of Physics, Cornell University, Ithaca, NY 14853, USA}

\date{10.07.2012}

\maketitle


\begin{abstract}

The interplay of bulk and boundary scattering is explored in a regime where quantum size effects modify mesoscopic transport in a degenerate Fermi liquid film of $^3$He on a rough surface. We discuss mass transport and the momentum relaxation time of the film in a torsional oscillator geometry within the framework of a quasiclassical theory that includes the experimentally determined power spectrum of the rough surface. The theory explains the anomalous temperature dependence of the relaxation rate observed experimentally. We model further studies on $^3$He confined in nanofluidic sample chambers with lithographically defined surface roughness. The improved understanding of surface roughness scattering can be extended to the analogous system of electrons in metals and suggests routes to improve the conductivity of thin metallic films. 

\end{abstract}

\section{Introduction}
Thin metallic nanowires are ubiquitous as electrical interconnects in modern-day devices. The performance of these wires given by their resistivity is limited by surface scattering. Quantum size effects combined with the effects of substrate roughness give rise to a mixing of surface scattering and inelastic scattering channels in such systems of confined degenerate Fermi liquid. Mattheissen's rule is violated and transport properties are tuned by the nature and extent of roughness.  In this paper, quantum transport is studied in a model system of a $^3$He film confined on a surface with roughness, whose power spectrum is experimentally determined. A theoretical treatment of the quantum size effects with the constraint imposed by the rough boundary is attempted here within a quasiclassical formalism.

The effect of a boundary is usually incorporated as a boundary condition in the theory for bulk transport \cite{prl2,prl3}. The  physical characteristics of the boundary surface are parametrized in terms of phenomenological surface parameters, such as a surface specularity coefficient. An alternative approach first suggested by Tesanovic {\it{et. al.}} \cite{Tesanovic} and independently by Trivedi {\it{et. al}}  \cite{Trivedi} maps the problem of a film with rough boundaries to an equivalent problem with flat boundaries and an added disorder potential in the confined bulk. This virtual disorder provides an additional scattering channel for quasiparticles which also scatter inelastically via binary quasiparticle scattering processes in the bulk.  For temperatures below $T\sim 100$ mK ($T\ll T_F$) , liquid $^3$He is well described by Fermi liquid theory : the inelastic scattering rate, $\tau_{in}^{-1} \propto T^2$  with inelastic mean free path $\lambda_{in} = 65/T^2$ ( $\mu$m  mK$^2$). The rate for elastic scattering off the virtual disorder potential arising from roughness depends on the power spectrum of the surface roughness. The effective scattering rate has an anomalous temperature dependence in a regime where  $\lambda_{in}\ll k_F^{-1} (k_F R)^2$ where R is the correlation length of surface inhomogeneities for a surface with a Gaussian form of the roughness power spectrum. We discuss this anomalous behaviour in the context of experiments performed at Royal Holloway University of London \cite{Casey2004} (RHUL) and formulate a theory for transport in these systems.

A brief outline of the experiment in Section 2 is followed by a theoretical development of the mapping transformation and the effect of confinement on the effective scattering rate in Section 3. The theory is used to predict mass and momentum transport in cavities with engineered surface roughness in Section 4.

\section{Experiment}

The relaxation rate of liquid $^3$He in thin films of thickness in the range $100-300$ nm was measured in a torsional oscillator, details of which are described elsewhere \cite{PRL}. The films were deposited on polished silver surfaces, whose surface roughness power spectrum was determined by atomic force microscopy (AFM) scans. The film was observed to decouple from the substrate even in the normal (Fermi liquid) state and the relaxation rate was observed to be linear with $\omega\tau_{osc}\sim 1$ over a range of temperature $10-100$mK.  The AFM scans revealed a featureless surface, details of which are reported elsewhere \cite{PRL}. Decorating the polished silver surface with big silver particles that contributed to additional surface scattering destroyed this anomalous effect and resulted in full coupling between the film and the substrate. The theoretical approach discussed below explains this film slip in terms of surface scattering from a rough surface, the power spectrum of which is particularly determined.

\section{Theory}

Consider a $^3$He film of thickness $L$. For quasiparticles with momentum $k_F^{-1} \ll L$,  the motion transverse to the film is quantized with $k_{F,z} = \pi n/L$ and the quasiparticle energies are quantized in a set of minibands $\epsilon_n = n^2 \pi^2 \hbar^2/ 2mL^2 + q^2/2m$, where $q$ is the two-dimensional quasiparticle momentum in the plane of the film. The inelastic binary quasiparticle scattering amplitude, $w$, is given in terms of Fermi liquid parameters and the phase space for scattering quasiparticles. The inelastic scattering rate $\tau_{in}^{-1} \propto N_F k_B T \cdot w \cdot N_F (k_B T/\epsilon_F) \propto T^2$ with $N_F$ being the quasiparticle density of states at the Fermi level,. Hence, for quasiparticles in a  slab of $^3$He with smooth walls,  the inelastic scattering rate $(\tau_{in}^{-1})_{smooth-slab} \propto T^2$, which is the quadratic temperature dependence also known in bulk $^3$He. Now consider a rough substrate surface with a height profile given by $\xi(x,y)$. We let the average height over the surface $\langle \xi\rangle = 0$ where $<...>$ denotes averaging over the surface. For $\xi \ll L$, the minibands have energies given by
\begin{equation}
\epsilon_n(x,y) = \frac {n(x,y)^2 \pi^2\hbar^2}{2mL^2 (1-\xi(x,y)/L)^2} + \frac{q^2}{2m}\,\,\,.
\end{equation}
The phase space available to scattered quasiparticles is now modified by roughness. To $\vartheta ((\xi/L)^2)$ (remember $\langle \xi\rangle = 0$),  $(\tau_{in}^{-1})_{rough-slab} \propto T\langle \xi \xi' \rangle$ where $\langle \xi \xi' \rangle \equiv \langle \xi(x,y) \xi(x',y') \rangle_{(x',y')}$ is the surface roughness structure factor. By a phase space consideration, we argue that the inelastic scattering rate has a linear temperature in a film with a rough substrate.

Confinement quantizes the quasiparticle momenta in the transverse direction, with the quasiparticle energy minibands being local as a consequence of the (local) roughness. Quasiparticles undergoing binary inelastic scattering scatter into a modified phase space for final states, determined locally by confinement and roughness. The exact form of the structure factor of the surface roughness can be incorporated into a theoretical formalism first developed by Meyerovich \cite{Mey98,Mey02} and coauthors.

The Hamiltonian for quasiparticles with momentum, $\vec{p}$ in bulk $^3$He is $\hat{H_0} = p^2/2m$. When confined to a slab of thickness, $L$, we impose walls at $z=\pm L/2$. For a film with one free surface and another surface with height profile, $\xi(x,y)$, the walls are at $z = L/2 - \xi(x,y)$ and $z = -L/2$.  A coordinate transformation $Z = (z+\xi/2)/(1-\xi/L)$ \cite{Tesanovic,Trivedi} maps this system to one with flat walls at $Z = \pm L/2$ with an added disorder potential in the bulk, $\hat{H} = \hat{H_0} + \hat{V}$ with $\hat{V} = (\xi/mL) P_Z^2$. Details of the transformation procedure have been discussed by Meyerovich and coauthors \cite{Mey98,Mey02}. A perturbation theory can then be carried out for elastic scattering via the disorder potential $\hat{V}$. Surface averaging leaves only those terms in the self-energy that are even in $\xi/L$ as the odd terms vanish (since $\langle \xi\rangle =0$). The leading term in the self-energy is proportional to the Fourier transform of the structure factor viz., the surface roughness power spectrum, $\zeta(\vec{q}-\vec{q'}) \equiv \langle \mid\xi(q)\mid^2\rangle $ where $\vec{q}$ is the Fourier transform of the two-dimensional vector $(x,y)$ in the plane of the surface with roughness profile $\xi(x,y)$.. The effective relaxation rate for a $^3$He film has been calculated by Meyerovich \cite{Mey01}. Bowley {\it{et. al.}} \cite{Bowley} calculated the momentum relaxation rate in a torsional oscillator geometry and found a linear dependence on temperature.

For films used in the RHUL experiments \cite{Casey2004}, the surface roughness power spectrum $\zeta$ was determined by high-resolution AFM scans. We have applied the formalism described above to derive an expression for the relaxation rate in a torsional oscillator, in terms of the {\it{measured}} power spectrum. We start from the expression for the effective quasiparticle relaxation rate derived by Meyerovich \cite{Mey98},\cite{Mey01},\cite{Mey02}, include the momentum transferred by the motion of the torsional oscillator and calculate the oscillator relaxation rate $\tau^{-1}_{osc}$. A change of variables is easily done then and $\tau^{-1}_{osc}$ can be expressed in terms of the {\it{measured}} power spectrum $\zeta(\vec{q}-\vec{q'})$.
\begin{equation}
\label{tauosc}
\frac{1}{\tau_{osc}} = \frac{3\sqrt{2}}{32\pi} \frac{1}{\tau_{in}}(\frac{k_F}{L})\sqrt{k_F\lambda_{in}}  \int _S\zeta(\vec{q}-\vec{q'}) \frac{t}{s}\sqrt{1-t^2}(t^2 -2ts+1)
\end{equation}
with
\begin{equation}
t = \frac{u}{k_F} + \sqrt{1 - \frac{v^2}{k_F^2}} \,\,\,;\,\,\, s = \sqrt{1 - \frac{v^2}{k_F^2}}\,\,\,\,\,,
\end{equation} 
where the integral is over the area of the substrate surface and $(\vec{q}-\vec{q'}) = (u,v)$ :  $u$ and $v$ are two-dimensional vectors in momentum space that span the rough surface, $S$ to obtain the power spectrum.

\section{Results}

\begin{figure}
\begin{center}
\vspace*{0.7cm}
\includegraphics[width=0.75\linewidth,keepaspectratio]{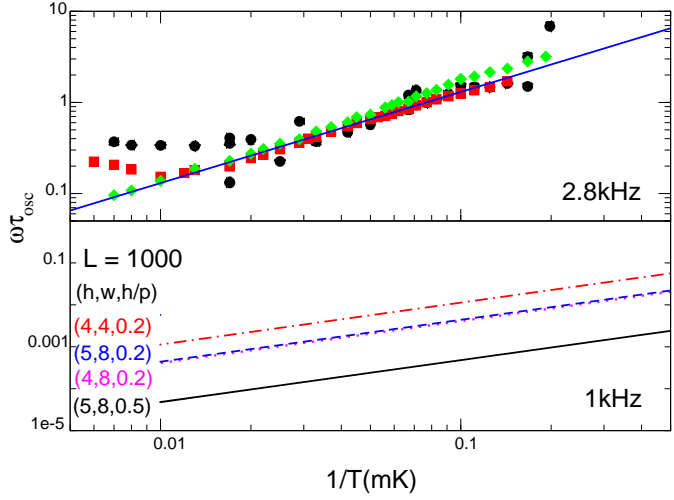}
\end{center}
\caption{(Color online) Top : The measured values of $\omega\tau_{osc}$ as a function of inverse temperature showing the linear temperature dependence for films of thickness $100-300$nm \cite{Casey2004} in a torsional oscillator operating at $2.8$kHz. The solid line shows the calculated value with $L$ as fitting parameter. Good fit obtained for $L=150$nm. Bottom : Predicted values for various surface roughness profiles engineered as an array of posts of height $h$, width $w$ and periodicity $p$ (in nms). Film thickness $L=1000$nm.}
\label{Fig3}
\end{figure}

The calculated relaxation rate $\tau_{osc}^{-1}$ is linear in temperature and shows good quantitative agreement with experimental data as shown in Fig.1.  The fitting parameter to the data in Fig.1 is the film thickness (fit to $L=150$nm) which is in the right ballpark for the films used in the experiment ($100-300$nm). The mixing of elastic scattering off the rough surface and inelastic binary quasiparticle scattering, including the exact details of the power spectrum for roughness, is seen to explain the anomalous film decoupling observed in such systems in the regime where weak elastic processes mix with bulk scattering. 

As the thickness of the film increases, and/or the temperature increases, the quasiparticle spectrum approaches a continuum and the scattering approaches the bulk limit. The relaxation rate recovers the bulk quadratic temperature dependence. Using the estimates of roughness of the order of that obtained experimentally (typical Gaussian power spectrum fit with average height $\sim 10$nm and correlation length $\sim 100$nm), this crossover to the bulk limit occurs at temperature $> 100$mK for films of thickness $L\sim 300$nm. Or, at low temperatures $T\sim 10$mK, this crossover occurs for films of thickness $L > 2500$nm. On the other hand, for smoother surfaces and decreasing roughness, the quasiparticles become insensitive to the roughness, leaving only the effect of confinement when $k_BT/\epsilon_F\sim\zeta/L^2$.

In general, this effect may be observed in Fermi liquid systems at temperature $T\ll T_F$ when the elastic and inelastic scattering rates are of comparable magnitudes. At low temperatures, thermal quasiparticles reside in an energy shell of size $k_B T \ll \epsilon_F$. The energy gap between quantized minibands  is related to the roughness and can be estimated to be of order $\sim \vartheta(1)\cdot \pi^2\hbar^2/2mL^2$.  The quantum size effect is observable when the gap between minibands is comparable to the inelastic scattering rate $\Delta\epsilon_{bands}\sim\tau_{in}^{-1}\le k_BT\ll \epsilon_F$ . The ratio of the strength of both scattering channels is $\Delta\epsilon_{bands}/(\hbar/\tau_{in}) = (1/k_FL)(\lambda_{in}/L)$. For the case of films of liquid $^3$He, the inelastic mean free path $\lambda_{in}=65\mu$m /T(mK)$^2$, $\epsilon_F \sim 1$K. For mK temperatures, $k_BT\ll\epsilon_F$ and $k_F L\gg 1$. We get $\Delta\epsilon_{bands}/(\hbar/\tau_{in}^{-1})\le 1$ for films of thickness $L\sim 300$nm, as observed in the experiments. The interband scattering is weak compared to the bulk scattering and momentum transfer in the transverse direction is ineffective. This is the regime in which the oscillator relaxation time is linear in temperature.

A direct analogy may be made to the case of electrons in metal thin films. Inelastic mean free paths in this case, $\lambda_{in} \le 1$nm and $\tau_{in} \sim 10^{-14}-10^{-15}$sec at room temperature, $T\ll T_F \sim 1000$K. For films of thickness $L\le 10$nm, the ratio $\Delta\epsilon_{bands}/(\hbar/\tau_{in}^{-1})\le 1$. Here again, interband scattering is weak compared to the bulk scattering and momentum transfer in the transverse direction is ineffective. For the case of Gaussian surface roughness power spectrum, the relaxation time is linear in temperature for roughness with correlations lengths, $R\ge 10$nm. This effect can hence be observed in thin ($\le 10$nm) metal films of this roughness and can be useful in engineering conductivities in such films.

Recent advances in the fabrication of nanofluidic cavities allow the study of helium films of well-determined thickness with fully characterizable surfaces\cite{CornellcavitiesRevSci}. The surfaces can be textured to engineer a target surface roughness that can be modelled within the theory for roughness in the limit, $\xi\ll L$. Consider a model engineered sample surface composed of a square lattice of rectangular posts of height $h$, width $w$ and periodicity $p$. The surface roughness power spectrum can be calculated and equation (\ref{tauosc}) used to compute the oscillator relaxation rate. The predicted rate for a range of roughness parameters is shown in Fig.1. Such a calculation can be used to fabricate cavities that couple helium over a given range of experimental parameters. This is an extremely useful tool that can be used to prepare cavities for experiments on superfluid $^3$He. It also has potential implications for the engineering of high quality metallic nanoscale interconnects in electronic devices.

\section{Conclusions}
The effect of confinement and the interplay of the coexistence of bulk and boundary scattering can be observed in Fermi liquid films in a regime, $k_F L\ll 1$. Both scattering channels are important over a range of temperature and film thicknesses. An anomalous linear temperature dependence of the relaxation rate of $^3$He films is observed in a torsional oscillator. This effect can be explained by the mixing of a weak elastic scattering channel, that arises from the surface roughness, with an inelastic bulk scattering channel. An analogy can be made to thin metallic films in which the electronic conductivity can have an anomalous temperature dependence in this regime.

\begin{acknowledgements}
The authors would like to thank EPSRC Grants No. GR/ S20677 and No. EP/E054129, European Microkelvin Consortium (FP7 grant agreement no:228464), the National Science Foundation DMR-0806629 and the Leverhulme Trust.
\end{acknowledgements}



\end{document}